*Original Article*

# SmartSecChain-SDN: A Blockchain-Integrated Intelligent Framework for Secure and Efficient Software-Defined Networks

Azhar Hussain Mozumder[1], M. John Basha[2], Chayapathi A. R.[3]

*[1]Department of Computer Science and Engineering (AI&ML), CMR University, Bengaluru, Karnataka, India*
*[2] Department of Computer Science and Engineering, JAIN (Deemed-to-be University), Bengaluru, Karnataka, India*
*[3] Department of Information Science and Engineering, JAIN (Deemed-to-be University), Bengaluru, Karnataka, India*

*[1]Corresponding Author : azhar.mozumder2@gmail.com*

*Abstract - With more and more existing networks being transformed to Software-Defined Networking (SDN), they need to be more secure and demand smarter ways of traffic control. This work, SmartSecChain-SDN, is a platform that combines machine learning based intrusion detection, blockchain-based storage of logs, and application-awareness-based priority in SDN networks. To detect network intrusions in a real-time, precision and low-false positives setup, the framework utilizes the application of advanced machine learning algorithms, namely Random Forest, XGBoost, CatBoost, and CNN-BiLSTM. SmartSecChain-SDN is based on the Hyperledger Fabric, which is a permissioned blockchain technology, to provide secure, scalable, and privacy-preserving storage and, thus, guarantee that the Intrusion Detection System (IDS) records cannot be altered and can be analyzed comprehensively. The system also has Quality of Service (QoS) rules and traffic shaping based on applications, which enables prioritization of critical services, such as VoIP, video conferencing, and business applications, as well as de-prioritization of non-essential traffic, such as downloads and updates. Mininet can simulate real-time SDN scenarios because it is used to prototype whole architectures. It is also compatible with controllers OpenDaylight and Ryu. It has tested the framework using the InSDN dataset and proved that it can identify different kinds of cyberattacks and handle bandwidth allocation efficiently under circumstances of resource constraints. SmartSecChain-SDN comprehensively addresses SDN system protection, securing and enhancing. The proposed study offers an innovative, extensible way to improve cybersecurity, regulatory compliance, and the administration of next-generation programmable networks.*

## 1. Introduction

The proliferation of internet-enabled devices, cloud-native services, and high-speed communications necessitates the ongoing development of novel network topologies [1]. Conventional network architectures, which are mostly hardware-driven and statically configured, are finding it increasingly challenging to meet the growing needs for scalability, flexibility, and real-time performance [2]. SDN has revolutionised this sector by decoupling the data plane from the control plane, paving the way for new programmable components, easier policy enforcement, and centralized management [3]. With SDN, managers programmatically govern network behavior through a centralized software controller. It is made possible by the separation of both the control logic and forwarding processes [4]. The design offers several substantial benefits, including simplified networks, increased agility, and optimal resource utilization. The new and substantial vulnerabilities presented by SDN's centralized architecture outweigh these benefits [5]. The control plane, housed in a logically centralized controller, is the primary target of any attacker serious about compromising the network's availability, integrity, or confidentiality [6]. These vulnerabilities enable attackers to disable critical network services, intercept or divert traffic, and modify flow rules. Two primary communication modes exist in Software-Defined Networking (SDN) systems for data and control channels, referred to as in-band and out-of-band, respectively [7]. Unlike in-band configurations, which allocate separate communication channels, in-band mode permits control and data traffic to use the same network connections [8]. Although in-band installation is quicker, the control plane is more vulnerable to data plane attacks. Hitchhiking on a controller grants an attacker complete control of the network if they manage to gain access to a switch [9]. Thus, protecting the SDN controller and its interfaces for communication is crucial for maintaining a secure and stable network [10]. It is possible to utilize SDN's programmable interfaces, which incorporate control logic. Unsecure authentication techniques are exploited by hostile actors through the introduction of unapproved OpenFlow rules or man-in-the-middle attacks [11]. The centralized controller cannot necessarily be sure of the switching devices it operates, a massive liability within the modern network environment, like 5G. The reliability of the OpenFlow devices can be tested in real-time using a subjective logic-based method. This allows the SDN controller to make contextual and adaptive decisions with respect to network security policies and the delivery of network services. The framework thus enables an active evaluation of the trust that can be executed to combat the inert quality of security models to foster network resilience to cyber-attacks [12]. As the number of SDN installations increases, traditional security measures developed for fixed, hardware-bound networks lose their efficacy in addressing these dynamic threats [13]. So, to protect SDN infrastructures, new approaches are required that integrate intelligent threat detection with tamper-resistant logging and dynamic traffic prioritization [14].

Progress in Artificial Intelligence (AI), particularly ML and DL, has aided adaptive network security lately. These systems use historical traffic data as a teaching tool to spot complex and surprising attack patterns. Unlike older, rule-based intrusion detection systems, those powered by machine learning and deep learning can quickly adapt to emerging threats [15]. Most people are interested in blockchain technology because it provides transparent, immutable, and decentralized data storage. Conducting forensic investigations and compliance audits is easier with the use of distributed ledger technology (blockchain), which ensures that critical network logs and intrusion detection system alerts cannot be altered [16]. Blockchain and AI have converged, creating an exciting opportunity to reassess security frameworks for SDN systems.

While this has some potential, existing research highlights several significant limitations. Currently, the majority of Intrusion Detection System (IDS) implementations rely on shallow learning techniques, which are inadequate for safeguarding SDN systems against more complex attacks [17]. Problems arise when solutions rely on out-of-date or insufficiently diverse data, which limits their practicality. Registering devices, authenticating users, and connecting with controllers were the primary uses of blockchain integration with SDN till now. Utilizing blockchain technology in a limited number of systems makes it possible to guarantee the legitimacy and veracity of signals from intrusion detection systems and other network events [18]. When bandwidth is limited, application-aware traffic prioritization usually has to take a back seat to other concerns, including Quality of Service (QoS). Modern networks enable a plethora of applications, including Voice over IP (VoIP), real-time video conferencing, background updates, and giant data transfers. Managing these services without intelligent traffic classification leads to performance degradation and user dissatisfaction [19].

To seal the loopholes and make SDN deployments safer, more dependable, and more efficient, this article introduces SmartSecChain-SDN, an intelligent security architecture that incorporates blockchain technology. All three components work together to form the proposed architecture: (1) ML/DL-based intelligent intrusion detection; (2) blockchain-protected auditable log storage; and (3) application-aware traffic prioritization via quality-of-service enforcement. Interconnected in an architectural fashion, these components form a whole that can manage responsibility, performance, and detection. With the aid of various deep learning and machine learning models, SmartSecChain-SDN enhances intrusion detection accuracy while reducing false positive rates. Among these models are CatBoost, eXtreme Gradient Boosting (XGBoost), Random Forest (RF), and a CNN-BiLSTM hybrid implementation. The InSDN dataset, a massive archive of data assembled solely for the purpose of evaluating SDN security protocols, was utilized to train a multi-model ensemble. With a hybrid method that combines standard ensemble models with deep sequence learners, SmartSecChain-SDN can detect any attack, whether it is a large volumetric assault or a stealthy infiltration. Typical detection systems struggle to spot these types of attacks.

By leveraging Hyperledger Fabric, a permissioned blockchain platform, SmartSecChain-SDN ensures that security logs are immutable, intact, and auditable. One advantage of Hyperledger Fabric over public blockchains is its modular design, which allows for fast throughput, transactions that safeguard user anonymity, and other similar features. By recording intrusion detection system alerts and security events on the blockchain, the technology produces reliable forensic investigations and satisfies regulatory compliance requirements. This leaves the evidence untouchable by any one party. To further reduce computing cost and maximize scalability, a permissioned blockchain is utilized to include only reliable companies in the consensus procedure. The network's administrators, controllers, or security agents could fall into this category. Key to SmartSecChain-SDN is intelligent, application-based traffic management. This component manages the real-time classification of traffic based on application type, criticality, and Service-Level Agreements (SLAs). Priority services, such as VoIP and video conferencing, receive bandwidth allocation priority during peak demand periods, while software upgrades are given lower priority. Optimized network performance and user experience are ensured by this dynamic QoS policy enforcement in instances where bandwidth is constrained or conflicting demands for traffic arise. Intelligent flow control is crucial in commercial, industrial, and Internet of Things (IoT) settings. After extensive testing with the Mininet emulator, OpenDaylight, and Ryu SDN controllers, the viability and utility of SmartSecChain-SDN were ascertained.

In a virtual environment that matches real-life SDN conditions, we test the machine learning models, blockchain infrastructure, and quality of service management modules. Utilizing the InSDN dataset, it is possible to evaluate the system's performance in several domains, including intrusion detection accuracy, false positive rate, blockchain transaction latency, quality of service compliance, and bandwidth utilization. Evidence from this study proves that SmartSecChain-SDN offers unmatched performance and security compared to more traditional solutions.

The objective is to develop a robust, flexible, and intelligent framework that can protect SDN environments from a wide range of cyber risks and enhance their operational efficiency. This research aims to design, develop, and deploy a multi-model machine learning pipeline capable of detecting diverse categories of real-time intrusions within Software-Defined Networking (SDN) environments. The proposed architecture, SmartSecChain-SDN, integrates advanced AI-driven intrusion detection with blockchain-based immutable logging and intelligent Quality of Service (QoS) management to address existing gaps in SDN security frameworks. In this system, multiple machine learning and deep learning models will operate in a parallel or hybrid configuration to classify threats ranging from Denial of Service (DoS) and Man-in-the-Middle (MitM) attacks to more sophisticated Advanced Persistent Threats (APTs). Detection decisions and anomaly scores will be securely recorded on a Hyperledger Fabric blockchain, ensuring tamper-proof logging, traceability, and accountability for security events. To guarantee service continuity and performance assurance, a novel application-aware QoS framework will be embedded, capable of real-time traffic prioritization based on application criticality, service-level agreements, and dynamic network conditions. This approach ensures that essential services such as industrial control systems or real-time IoT applications maintain high performance even under attack scenarios. The architecture will be implemented and validated on a virtual SDN testbed integrating OpenDaylight and Ryu controllers, the Mininet network emulator, and the InSDN dataset to simulate realistic attack patterns and traffic flows. By combining AI for threat detection, blockchain for secure logging, and QoS-driven traffic orchestration in a unified system, SmartSecChain-SDN offers a scalable, programmable, and security-aware SDN control solution that is robust against both conventional and emerging threats. This research represents one of the first comprehensive attempts to jointly optimize security, integrity, and service quality in next-generation programmable networks, providing a blueprint for future cybersecurity-aware SDN control architectures in industrial, enterprise, and critical infrastructure deployments. The main contributions of the paper include

- The hybrid Intrusion Detection System (IDS) engine uses deep learning frameworks and ensemble machine learning models to produce low false positive rates and high detection accuracy in SDN environments.

- A Hyperledger Fabric blockchain module that securely and unchangeably records security occurrences, making forensic investigation and audit compliance easier.
- The service administration system dynamically prioritizes network traffic based on applications and service criticality. Service level agreements are met while maintaining efficiency.
- Comprehensive experimental verification utilizing the Mininet simulator, OpenDaylight and Ryu controllers, and the InSDN dataset proves the system's superiority over baseline techniques.

## 2. Literature Survey

Ali et al. [20] introduce the Federated Learning-Enhanced Blockchain (FL-BCID) architecture for privacy-protecting intrusion detection in Second IoT situations. Federated Learning (FL) and blockchain technology ensure decentralized model training, data consistency, trust, and tamper resistance across all IIoT nodes in the architecture. FL is used to build a lightweight intrusion detection model to secure sensitive data. This model was trained cooperatively on edge devices: smart contract-enabled blockchain systems record model changes and anomaly ratings for accountability. The system beat baseline centralized systems in ToN-IoT and N-BaIoT experiments. Communication overhead was reduced by 41% and accuracy was 97.3%. This technique offers privacy, scalability, and resilience, which are crucial for ensuring safe industrial operations. A promising alternative to existing IoT security concepts is the FL-BCID system.

Alqahtani et al. [21] observed that Advanced Persistent Threats (APTs) that are stealthy and adaptable are good at avoiding detection. Concept drift occurs when the statistical features of input data change over time, particularly concerning how attackers behave. It was demonstrated to be a significant issue for utilizing machine learning to enhance the accuracy of Intrusion Detection Systems (IDSs). The goal is to enhance IDS accuracy in identifying threats by developing an incremental, hybrid, adaptive Network Intrusion Detection System (NIDS) as part of the research. Tests on several datasets indicated that the model can detect stealthy attacks in SDN networks. The model identifies idea drift to maintain performance in changing situations.

Jamshidi et al. [22] investigated how Software-Defined Networking (SDN) impacts Machine Learning (ML)- based Intrusion Detection Systems (IDS) deployed at the edge of Internet of Things (IoT) infrastructures. The study found that deep learning techniques have smaller resource overheads than machine learning-based intrusion detection systems in response to real-time cyber threats. SDN's centralized control improved resource management, but increased overhead when risks were present. An Analysis of Variance (ANOVA) supports the findings, which reveal trade-offs between edge-based Internet of Things detection accuracy and system performance. For SDN-based smart infrastructures, Mustafa et al. [23] suggested C-RADAR, a centralized anomaly detection and response platform. Self-attention and LSTM networks in a deep learning architecture enable the machine to grasp contextual and temporal patterns in network traffic. The SDN controller detects and responds to rising risks in real time with C-RADAR. The model is integrated with the SDN controller for real-time detection and response to intrusions. The framework detects several attack types with high accuracy and low false positives. The centralized structure generates scalability and latency issues in large-scale or high-throughput network deployments.

Núñez-Gómez et al. [24] claim that S-HIDRA is a system that uses SDN and blockchain technology to control containerized services in fog computing. Smart contracts on the blockchain enable the orchestration of tasks in a decentralized and immutable manner. SDN allows networks to be reconfigured quickly and easily to accommodate changing workloads on fog nodes. This solves the problems with centralized cloud architectures. The architecture employs a domain-based approach to address fog nodes' geographical spread and movement, thereby delivering low latency and high service availability. A proof-of-concept implementation demonstrated that S-HIDRA is effective, although resource orchestration is more crucial than traffic-level security enforcement or direct intrusion detection.

The work of Commey et al. [25] is concerned with cybersecurity issues regarding Blockchain-based IoT (BIoT) systems. It creates an AI-driven Honeypot deployment model that is combined with an Intrusion Detection System (IDS) and smart contracts on IoT nodes. The model provides the ability to convert regular nodes into decoys upon the detection of suspect activity, thus ensuring higher network security. A game-theoretic analysis of the strategic interactions involving the attackers and the AI-enhanced IDS is conducted through a game-theoretic model, namely the Bayesian games. According to the study, emphasis will be on knowledge and prediction of complex attacks, which might seem normal in the beginning. One does not have to be a security expert to see the benefits of the proposed model compared to the conventional security methods based on topics that dynamically disintegrate the threats and, based on the smart contracts, provide automated and fast reactions. Simulations were used to compare the performance effectiveness of the honeypot deployment strategies. The results indicated the capacity of the model to achieve optimum security and efficiency of the operation as well. By drawing such conclusions, the authors demonstrate that the proposed approach can be used to establish the foundation for further developing more intelligent and dynamic defense mechanisms in BIoT systems.

Hyder et al. [26] demonstrated that a Software-Defined Networking (SDN) implementation of Moving Target Defense (MTD) can be utilized to counter Crossfire-style Distributed Denial of Service (DDoS) attacks. The network dynamically shifts network paths by updating the open flow traffic rules and redirecting flows that could be attacked. These changes are orchestrated by an intent-based SDN controller to redirect the traffic to decoy nodes and mix things up with the attackers. Experiments demonstrate the enhanced resilience of networks and alleviated link overload in the presence of attacks. The authors, however, indicate that overhead comes up with frequent reconfiguration and might propagate routing instability.

Unlike previous studies, Poorazad et al. [27] provided a combined method that addresses IIoT and SDN security. Our objective is to identify and avoid security issues in SDN-based IIoT architectures. This technique improves this. By working together, both components improve application and network layer security. This system starts with a software-defined network application-based convolutional neural network-based Intrusion Detection System (IDS). A blockchain-based system is the second component. The proposed solution reduces rule and command injection attacks on IIoT layers using Software-Defined Networking (SDN). The proposed IDS can successfully classify binary and multiclass data.

Putra et al. [28] proposed a blockchain-based Collaborative Intelligence Detection System to improve Intrusion Detection Systems (IDSs). This paradigm allows CIDS node users to exchange intrusion warnings and detection criteria. Most blockchain-based CIDS approaches assume nodes are innately trustworthy, which is wrong. Most proposals overlook the need to verify nodes on a routine basis. This paper presents a decentralized CIDS that emphasizes node trust. The method utilizes CIDS nodes to communicate detection criteria and identify new intrusions. The design enables scalability by storing shared trustworthy detection criteria in a decentralized system and delivering the trust calculation to the blockchain. The solution is tested on a lab-scale testbed to demonstrate its feasibility and performance against Ethereum platform benchmarks.

Sarhan et al. [29] propose a hierarchical blockchain-based federated learning system for a secure and private collaborative IoT intrusion detection system. Machine learning-based intrusion detection systems should employ hierarchical federated learning. The learning process and company data are protected. The smart contract will ensure accuracy, while the secure blockchain will handle operations and transactions. We tested the intrusion detection system using a lot of IoT data. A safe, machine-learning-based intrusion detection system that can uncover many threats without compromising user data will be the result.

Benoudifa et al. [30] present a dynamic controller placement framework, based on MuZero reinforcement learning and smart contracts. The system measures latency, the volume of traffic, and the connection of devices to optimize the locations of controllers. Blockchain provides evidence of the non-tamperability of placement decisions by using smart contract logging. The OpenDaylight controller is used to implement the solution in the Mininet. Results demonstrate better latency and security rates in comparison to the rigid approach. Nonetheless, MuZero training and blockchain consensus have overheads, which restrict their application in the edge environment.

SDN centralizes control for agility but enlarges the attack surface (e.g., controller saturation, rule manipulation). Recent surveys map threats and countermeasures, emphasizing the need for intelligent, controller-aware IDS that operate at line rate with low overhead. Deep Learning (DL) has become dominant for SDN IDS due to superior spatiotemporal feature extraction versus classical ML, but challenges remain around class imbalance, concept drift, and deployment efficiency [31]. Recent SDN-focused studies report strong performance from CNN/LSTM/transformer hybrids, meta-heuristic tuning, and multi-head architectures, often evaluated on SDN-specific corpora. Still, most works optimize accuracy in isolation and underplay system integration (e.g., logging, policy, QoS) [32]. General legacy datasets (e.g., KDD'99) misrepresent SDN realities.

InSDN adds controller/data-plane attack coverage and is widely referenced 2020–2025 for reproducible SDN IDS evaluation. Curated surveys from 2024 consolidate research using InSDN and highlight remaining gaps (e.g., multi-controller scenarios, mixed benign/attack traffic dynamics). A Kaggle mirror facilitates experimentation [33]. FL reduces raw data movement and supports edge-level training that is useful for multi-domain SDN/IoT. Recent work advances FL-IDS model selection and resource awareness on constrained devices, while broader surveys chart progress in FL-IDS for IoT and IIoT.

However, most studies stop short of tying FL outputs into operational SDN control loops or immutable audit trails. Blockchain adds tamper-evident, verifiable logging and decentralized coordination. 2024–2025 studies and surveys propose blockchain-enhanced IDS, sometimes coupled with FL, to improve trust, data integrity, and collaborative detection, but typically focus on the logging or collaboration plane, not end-to-end SDN performance/QoS. Emerging work explores blockchain in SDN security functions (e.g., firewalls) and shows feasibility, yet overhead/scheduling issues persist [34].

There have been advancements, but no single framework integrates intelligent intrusion detection, immutable logging, adaptive traffic management, and quality of service assurance in SDN contexts. In contrast, they focus on authentication, detection, and orchestration. This difference underscores the need for a blockchain-integrated, AI-driven design, such as

SmartSecChain-SDN, that can identify threats, ensure log integrity, and prioritize valuable application traffic in programmable and dynamic network topologies. Despite significant advancements in intrusion detection systems leveraging federated learning, blockchain, and SDN, existing approaches largely address isolated aspects such as privacy-preserving model training, immutable logging, adaptive orchestration, or specific attack mitigation.

Current solutions often suffer from limitations, including reliance on centralized control, lack of real-time node trust verification, absence of integrated QoS management, and inadequate adaptability to evolving threats like concept drift. While some frameworks combine two technologies (e.g., blockchain with IDS, SDN with deep learning), no unified architecture holistically integrates intelligent intrusion detection, tamper-proof logging, adaptive traffic management, and QoS assurance in programmable SDN-enabled IIoT environments. This gap highlights the need for a blockchain-integrated, AI-driven framework to ensure threat detection accuracy, maintain system resilience, and prioritise critical traffic under dynamic and high-load network conditions.

## 3. Proposed Methodology
### 3.1. Dataset Description

The InSDN dataset is associated with testing intrusion detection solutions. This new dataset includes all the malicious and good attacks on SDN standard components and is shown in Table 1. Attacks such as DoS, DDoS, brute force, web application, exploitation, probing, and botnet attacks are some of the considerations of SDN. In addition, data traffic is related to the most widely used application services, such as HTTPS, HTTP, SSL, DNS, Email, FTP, SSH, etc. The InSDN dataset is taken as the reference, with which Software Defined Networking Intrusion Detection Systems (IDS) can be quantified. It has denial-of-service, distributed denial-of-service, brute-force, exploitation, probing, botnet and web-based attacks, and it records over 343,000 malicious and benign traffic events.

HTTPS, FTP, DNS, SSH, etc., are patterned after normal communication practices. The data that was employed was generated in a virtualized lab with four virtual PCs and Mininet hosts. Many types of traffic were captured, such as the internal and external threats. Metasploitable 2 and DVWA were used to attack containerized systems, while Wireshark was used to collect PCAP traffic. With CICFlowMeter, feature extraction was performed, after which the data was ready to be processed through the machine learning analysis. More than 80 flow-level statistics were stored in the form of CSV by the application. The data are realistic, properly labelled, and large enough to test intrusion detection systems using SDN-integrated systems (e.g., SmartSecChain-SDN). The data is also broken down into Normal, Metsplotable-2 and Ocean View System classes.

**Table 1. In the SDN dataset parameter and structure**

| Parameter | Description |
|---|---|
| Dataset Name | InSDN – Intrusion Detection Dataset for Software Defined Networks |
| Purpose | Evaluation of Machine Learning-based IDS in SDN environments |
| Traffic Types | Normal and Malicious (DoS, DDoS, Brute Force, Web Attacks, Exploits, Probe, Botnet) |
| Normal Services Covered | HTTPS, HTTP, SSL, DNS, Email, FTP, SSH |
| Total Instances | 343,939 records |
| Normal Records | 68,424 instances |
| Attack Records | 275,515 instances |
| Traffic Capture Format | PCAP files captured via Wireshark |
| Feature Extraction Tool | CICFlowMeter (by Canadian Institute for Cybersecurity) |
| No. of Features Extracted | Over 80 statistical features (e.g., Duration, Protocol, Byte Count, Flow IAT) |
| Flow Directionality | Bidirectional flows (first packet determines flow direction) |
| Labelling Method | Based on Source IP, Destination IP, and attack context |
| Virtual Machines Used | - Kali Linux (attacker)<br>- Ubuntu (ONOS Controller)<br>- Ubuntu (Mininet + OVS)<br>- Metasploitable 2 (vulnerable services) |
| Virtual Hosts (Vhosts) | - h1, h2 (malicious)<br>- h3 (benign user)<br>- h4 (web server) |
| Container Deployment | DVWA server deployed via Docker on the OVS machine |
| Data Groups | 1. Normal Group<br>2. Metsplotable-2 Group<br>3. OVS Group |
| Normal Group Size | 3.58 GB across 10 directories |
| Metsplotable-2 Group Size | 669 MB across 5 attack categories (DoS, DDoS, Exploit, Probe, Brute Force) |
| OVS Group Size | 1.21 GB across 6 attack types (Botnet, Brute Force, DoS, DDoS, Web, Probe) |
| Capture Points | 1. Target machine interface<br>2. SDN Controller interface |
| Output Format | CSV files containing intrusion statistical flows |

### 3.2. Overall Structure of SmartSecChain-SDN

Machine learning, blockchain technology, and QoS-aware traffic control are combined in the SmartSecChain-SDN technology in order to secure and improve SDNs. The data plane of the SDN is the core of the architecture and collects and prepares network flows to undergo real-time analytics and processing. Others employ a hybrid IDS with a high level of detection of known and zero-day threats and low false positives. The approach used to develop a scalable and reliable intrusion detection system is described in the proposed framework shown in Figure 1 as decentralized training of the models, preprocessing of the data, secure data management, and smart contract tool implementation. To succeed in intrusion detection by machine learning detection of threats on the network demands the extraction of network flow characteristics with reference to volume, statistics and time. The features of a malicious application flow are abnormal length or shortness of its duration, unlike a benign application flow. The flow duration is calculated based on the $T_{flow} = t_{last\,pac} - t_{first\,pac}$ And the average packet size is represented in Equation (1) with the byte rate. $R_{byte} = \frac{\sum_{i=1}^{N} S_i}{T_{flow}}$ Intruders can avoid detection by altering packet sizes or utilizing preset sizes. Analysis of average size shows such tendencies.

$$\bar{S}_{pkt} = \frac{1}{N} \sum_{i=1}^{N} S_i \quad (1)$$

The average packet time difference is one metric of stealth attack frequency, which is irregular or periodic. Inter-arrival time ($\tau_{avg}$) in Equation (2) helps identify timing abnormalities.

$$\tau_{avg} = \frac{1}{N_1} \sum_{i=1}^{N-1} (t_{i+1} - t_i) \quad (2)$$

Where $N_1$ Is the number of packets in the flow and $t_i$ Is the arrival time of the packet. Equation (3) port entropy computes the destination ports used in the flow with $p_i$ Probability of occurrence of port $I$ (incoming packet). Low-entropy communication patterns are more likely to be determined, while high-entropy port scanning or spread attacks are more likely to be arbitrary.

$$H(p) = -\sum_{i=1}^{n} p_i \log_2(p_i) \quad (3)$$

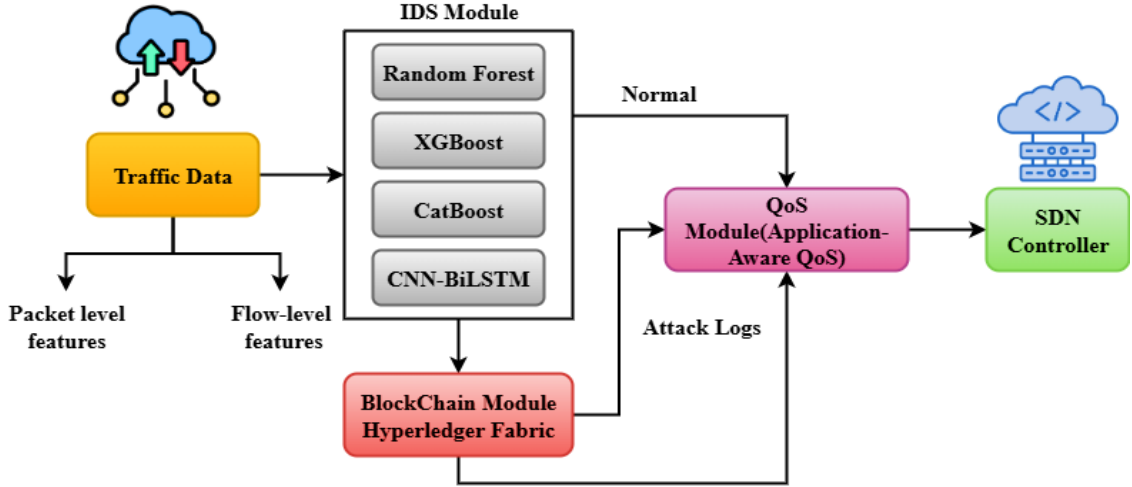

**Fig. 1 Structure of SmartSecChain-SDN**

Intelligent framework is an emerging architecture concept based on the exploitation of blockchain applications to overcome both efficiency and security issues of highly scalable networks, e.g., the Internet of Things (IoT). The design creates a robust and adaptable system using AI, Blockchain, and SDN. This framework integrates XGBoost, CatBoost and Random Forest and a CNN-BiLSTM deep learning model. Hyperledger Fabric is a permissioned blockchain that provides immutable storage of threats and system logs on top of secure and auditable logging and secure traceability of all IDS alerts and changes. Our smart contracts on the blockchain offer protection of the information and ease of operations. The access and decryption of data are only available to the parties that have been authorized in these contracts. The highest priority is data security.

An access control is a very important feature of a smart contract. Not only will it warrant safe access to all affected parties, but it will also limit the access of data retrieval and interaction to persons whose cryptographic keys are valid. The contract also offers an audit of all data transfers, a record of everything that has gone through, auditing and data provenance. Truth inspires credible information and instils confidence in collaborative learning. To share data, there is also an automated agreement carried out under smart contracts, making the management of data flow easier and unauthorized access cannot occur during machine learning and other multi-stakeholder inspired applications.

In severe instances, the control plane constrains network functionality with the help of application-sensitive quality of service norms. The constraints favour services with a low latency, such as VoIP and video conferencing and disadvantage traffic with less critical sensitivities, such as software updates and bulk downloads. Flow rules are implemented dynamically by an SDN controller (such as Ryu or OpenDaylight), depending on the level of security and the classification of the traffic.

Besides, adjustments to traffic and capacity according to network loads can be provided through the system. Mininet determines whether the framework would work by simulating its architecture and running it through security checks, intrusion detection and service quality and compliance in order to test how it would perform. SmartSecChain-SDN provides an intelligent, scalable, and unified solution for programmable digital infrastructures.

### 3.3. Intrusion Detection Model Design

Intrusion detection models detect unauthorized access to a computer system or a network. It is important to acquire the data, clean it, select characteristics, train the model, and evaluate it. The intrusion detection process shown in Figure 2 typically employs detection methods based on anomalies, signatures, or a combination of both. Such methods are employed to investigate the system activity and network traffic. Some of the methods are machine learning and deep learning. SmartSecChain-SDN can enhance threat detection in SDN environments by connecting CNN-BiLSTM, a sequential deep learning model, and common machine learning classifiers (Random Forest, XGBoost, and CatBoost). The hybrid method of rapid classification and context-aware sequence analysis can help in capturing more complex traffic patterns and, at the same time, minimize the false positives.

More than anything, the appropriate choice of traffic input improves learning, minimizes dimensionality, and eliminates superfluous variables. With CICFlowMeter, over 80 flow-based characteristics statistics were sampled in InSDN. Intradomain expertise and correlation results resulted in the 32 best qualities: protocol flags, flow entropy, packet Internal Address Translation (IAT), and byte rate. Since the difference in the scales may cause bias in the input, all the numerical features x were normalized through Z-score normalization as shown in Equation (4).

$$x_{norm} = \frac{x - \mu}{\sigma} \qquad (4)$$

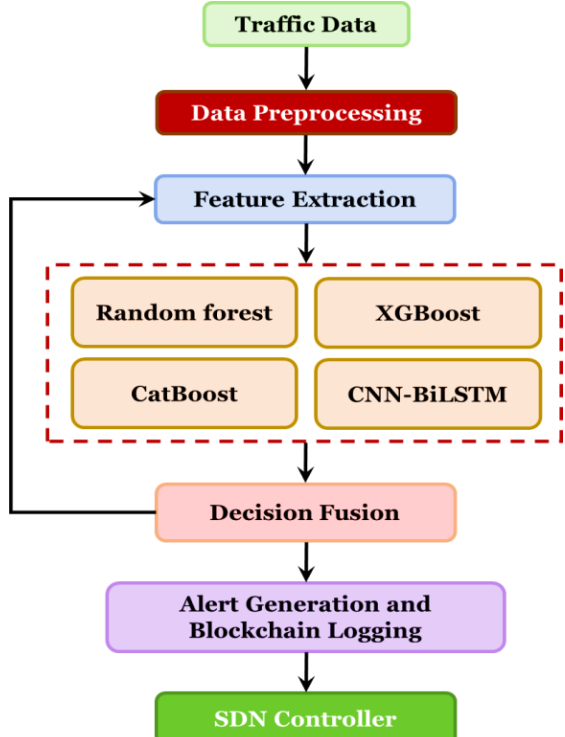

**Fig. 2 Intrusion detection workflow**

Where $y_i$ Is the actual target value of the i-th function, $\hat{y}_i^{t-1}$ denotes the predicted value, $f_t(x_i)$ Is the predicted value from the new model tree being added at iteration t? The next one is the CatBoost, which is a gradient boosting technique that makes category data work better based on the tree weight. $\alpha_k$ and the decision tree function $T_k(x_i$ Using ordered boosting and target statistics can help stop overfitting. The output is represented in Equation (7).

$$\hat{y}_c = \sum_{k=1}^{K} \alpha_k T_k(x_i) \tag{7}$$

Finally, CNN-BiLSTM, in which networks keep track of time-based connections, and CNN keeps track of space-based connections. Botnets and slow, covert attacks are two examples of flow patterns that depend on time and the output of the CNN shown in Equation (8).

$$f_j^{(l)} = \sigma\left(\sum_i x_i^{(l-1)} * w_{ij}^{(l)} + b_j^{(l)}\right) \tag{8}$$

And the LSTM hidden cell update is represented by Equation (9).

$$h_t = \tanh\left(W_f \cdot [h_{t-1}, x_t] + b_f\right) \tag{9}$$

Each of the models in the SmartSecChain-SDN ensemble has been hyperparameter-adjusted and trained extensively to be as effective as possible against SDN threats. In spite of the fact that Random Forest and gradient boosting models are more robust and simpler to understand, CNN-BiLSTM performs an excellent job of determining sequential patterns of attack. The CNN-BiLSTM model swiftly and precisely detects intrusions, even in highly volatile or adversarial network environments. In order to achieve even greater predictive accuracy and model resilience, the additional step of adopting a probabilistic ensemble voting mechanism ($\hat{\rho}$), illustrated by Equation (10), is commonly employed. The combination has the advantage of using the strengths of several different models and alleviating the poor generalization problem that can emerge with using only one classifier. Together, the ensemble brings out the merit of model diversity in creating a more effective and reliable IDS.

$$\hat{\rho} = \arg\max_{c \in C}\left(\sum_{m=1}^{M} w_m \cdot P_m(c \mid \vec{x})\right) \tag{10}$$

Where $w_m$ is the weight assigned to the m-th classifier, $P_m(c \mid \vec{x})$ denotes the probability output of the classifier for class $c$, given an input vector $\vec{x}$. The probabilistic voting algorithm means that classifiers with higher performance impact the decision to be established, so the intrusion detection system will gain more robustness and adaptability. Pseudocode 1 shows the step-by-step procedure of the proposed intrusion detection mechanism and the way to combine blockchain-based logging that will provide secure and immutable event recording.

The data was normalized and then divided into training and testing data, and missing values were replaced with mean substitution. One-hot encoding of categorical values was used. With the help of a multi-algorithmic deployment of deep learning networks and machine learning classifiers, a flexible and robust intrusion detection system is built. Each model works excellently with non-linearity, over-fitting, interaction of features and time varying patterns.

Stratified K-Fold Cross-Validation and optimization have been used to determine the best hyperparameters for each model so that it would provide the most accurate results and the minimum number of false positives. Each Random Forest decision tree is trained using a different set of data and attributes, which are selected randomly. A Random Forest is an ensemble of decision trees. It prevents overfitting and can handle noise with bootstrapping. Prediction for input $T_i(x)$ Equation (5) is based on the majority vote, where $n$ is the number of trees.

$$\hat{y}_{rf} = mode(\{T_i(x)\}_{i=1}^{n}) \tag{5}$$

Next, XGBoost uses regularization and a second-order Taylor approximation to optimize a loss function ($l$) and prevent overfitting, as shown in Equation (6). This approach is effective despite having limited data and a class imbalance.

$$\mathcal{L}(t) = \sum_{i=1}^{n} l(y_i, \hat{y}_i^{t-1} + f_t(x_i) + \phi(f_t)) \tag{6}$$

**Pseudocode 1: Intrusion Detection and Blockchain-Integrated Logging**

| |
|---|
| Inputs: Packet stream, Trained models, Flow Timeout ($\tau$), Blockchain API |
| Output: Intrusion Log $\leftarrow$ List of detected attack records stored immutably |
| 1. Initialize: |
|    Flow Table $\leftarrow$ {} |
|    Intrusion Log $\leftarrow$ [] |
|    Model Weights $\leftarrow$ {RF, XGB, CAT, CNN-BiLSTM} |
|    Voting Threshold $\leftarrow$ 0.5 |
| 2. For each incoming packet $p_i$ In PacketStream: |
|    flow_id $\leftarrow$ Compute FlowID($p_i$) |
|    Flow Table [flow_id].append($p_i$) |
|    Update $t_{last\,pac}$ for flow_id |
| 3. Periodically check expired flows: |
|    For each flow_id in the Flow Table: |
|       If Current Time $- t_{last\,pac} \geq \tau$: |
|       F $\leftarrow$ Extract Flow Features (Flow Table[flow_id]) |
|    Step 1: Compute Derived Statistical Features --- |
| $T_{flow} = t_{last\,pac} - t_{first\,pac}$ |
| $\bar{S}_{pkt} = \frac{1}{N}\sum_{i=1}^{N} S_i$ |
| $R_{\text{byte}} = \frac{\sum_{i=1}^{N} S_i}{T_{\text{flow}}}$ |
| $\tau_{avg} = \frac{1}{N_i}\sum_{i=1}^{N-1}(t_{i+1} - t_i)$ |
| $H(p) = -\sum_{i=1}^{n} p_i \log_2(p_i)$ |
| Step 2: Normalize Features ---Based on $x_{norm} = \frac{x-\mu}{\sigma}$ |
| Step 3: Perform Predictions with Each Model --- |
|    PredScores $\leftarrow$ {} |
|    For each model M in Trained Models: |
|       label $\leftarrow$ M.predict ($x_{norm}$) |
|       PredScores[label] += Model Weights[M] |
| Step 4: Weighted Decision Fusion --- |
|    final_label $\leftarrow$ Arg Max (Pred Scores) |
|    If final_label $\neq$ "Normal" AND PredScores[final_label] > VotingThreshold: |
|       Alert $\leftarrow$ { |
|         "FlowID": flow_id, |
|         "Label": final_label, |
|         "Confidence": PredScores[final_label], |
|         "Timestamp": Now() |
|       } |
| Step 5: Blockchain-secured Logging --- |
|       BlockchainAPI.submit_transaction(Alert) |
|       Intrusion Log.append(Alert) |
|    Delete Flow Table[flow_id]  // Clean processed flow |
| 4. Return IntrusionLog |
| END |

### 3.4. Blockchain Logging and Smart Contract Integration

SmartSecChain-SDN features an integrated Hyperledger Fabric blockchain that securely logs intrusion detection system alerts. This makes data more reliable, easier to trace, and harder to deny. With this approach, smart contracts, or chain codes, safely and permanently store warning data and conduct forensic audits. Alerts from the ensemble identification and threat system provide a timestamp, the expected type of attack, the Flow ID, and the level of confidence in the detection. When detection confidence exceeds a predetermined level $\theta$, a smart contract goes into effect right away. Before recording, this warning is delivered to the smart contract. The logging condition is enforced as in Equation (11):

$$if\ C_{score} \geq \theta \Rightarrow (Flow\ ID, Class, Confidence, Time)\quad(11)$$

Each alert transaction is digitally signed and hashed using SHA-256 to ensure integrity before submission, and digital signature generation is given in Equations (12) and (13):

$$H_{alert} =\ SHA\ 256\big(log_{data}\ \|\ timestamp\ \big)\quad(12)$$

$$Sig\ =\ sK_{priv}\ (H_{alert})\quad(13)$$

Where $sK_{priv}$ is the sender's private signing key. Hyperledger peer nodes send signed transactions to the ordering service. The ordering service then organizes blocks of peer transactions. They are then committed to the unchangeable electronic ledger. Under the consensus policy, the transaction will only be allowed if all necessary peers agree. This ensures only legitimate cautions are saved. Smart contracts analyze alarm formats and source identities, notify SDN management in real-time, and make the audit requests very simple through a RESTful interface. A recorded alarm can be recovered for forensics, policy review, or compliance audits. Once captured in the record, an alert becomes a part of the ledger permanently. Old warnings can be accessed via querying the blockchain with the appropriate Flow ID. The outcome of such a query contains vital alert metadata.

$$Query\ Alert_{flow\ ID} = \{\ Class, Confidence, Time, Action\}\quad(14)$$

Figure 3 presents the process scheme that the SmartSecChain-SDN architecture incorporates to store alerts on the blockchain. The chaincode (smart contracts) is triggered as soon as the algorithm identifies an abnormality with a confidence score of a previously agreed threshold. A node in the Hyperledger Fabric network acts as the host and performer of this contract. Before processing the alarm data, the smart contract performs a verification step to ensure that it is genuine and originated from a reputable IDS supplier. If the findings come back negative or the detection confidence is low, the record generation will not proceed according to the contract. To help system administrators investigate or fix the issue quickly, the contract saves the data and sends them a real-time alert. After a smart contract accepts a transaction, the ordering service, which is the entity that receives the funds, is responsible for ensuring that all the blockchain nodes reach a consensus. As part of its ordering process, the blockchain is responsible for creating new blocks and sequentially organizing transactions in chronological order. Now these blocks need to be added to the immutable ledger. It is difficult to alter any transaction that passes through the blockchain, as the data recorded on it is both immutable and fully auditable. The query feature in the system can permit a user authorized to view logged alerts, like the auditors or the network administrators, to further query the system about any information about the alerts. As you can see, the secure closed-loop system for detection, logging, and verification relies on query answers retrieved straight from the same immutable ledger.

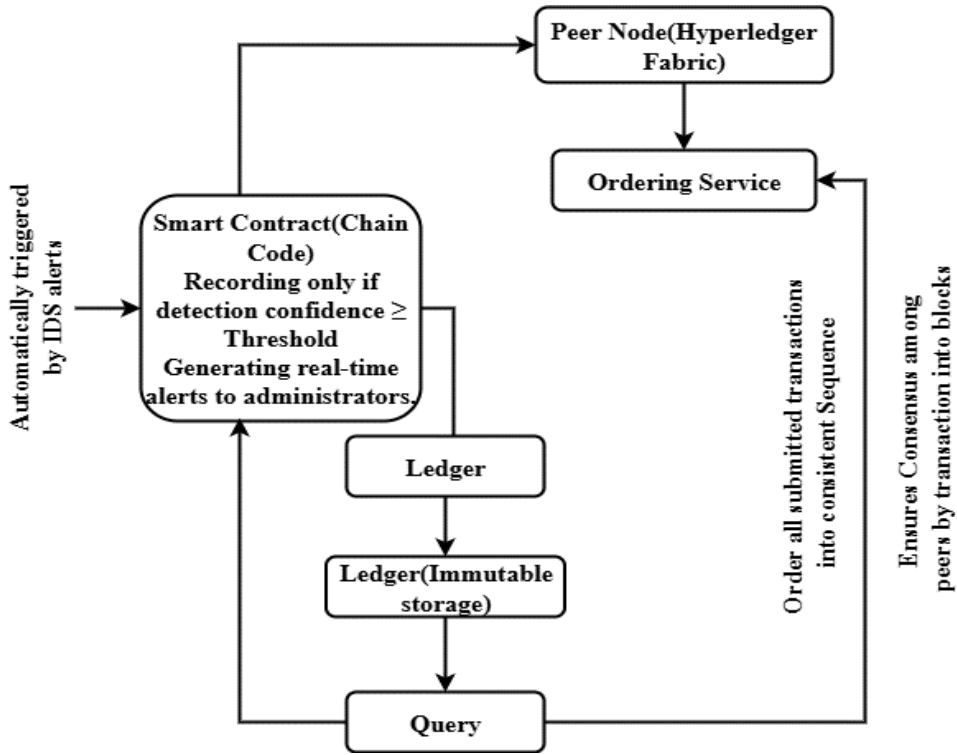

**Fig. 3 Blockchain logging and smart contract integration**

### 3.5. QOS Enforcement and SDN Controller Integration

The controller manages network resources and traffic flows, and it is therefore necessary to impose quality of service in SDN. When demands for different levels of services are made, the controller monitoring the network determines the optimal pathways and configures the necessary devices to follow the rules. The integration efficiently and dynamically controls network resources to ensure high-priority traffic receives latency and bandwidth guarantees. Figure 4 illustrates a detailed representation of the SmartSecChain-SDN framework's QoS enforcement procedure, integrated with the SDN controller.

Alerting the appropriate parties should be the initial action the intrusion detection system takes when it detects potentially dangerous network activity. This notice includes a timestamp, a detection confidence score, the type of attack, and the Flow ID, among other details. Immediately upon warning generation, it is sent on to the module responsible for handling warnings. At that location, we verify the information and collect data for the next phase of decision-making. The next step is to check the Alert-to-Policy Mapping Table. This table outlines the relationships between various network actions (such as dropping, rerouting to a honeypot, or prioritizing) and different types of attacks (such as DDoS, botnet, VoIP, or even benign traffic). Threat severity estimation based on traffic and alert factors is represented by Equation (15).

$$\mathcal{T}_{sev} = \alpha \cdot log\left(\frac{B_{src}}{B_{total}}\right) + \beta \cdot \mathcal{C}_{score} + \gamma \cdot R_{freq} + \delta \cdot H_{entropy}$$
(15)

Where $\frac{B_{src}}{B_{total}}$ Is the bandwidth usage of the suspicious source, $\mathcal{C}_{score}$ is the ML model's confidence score, $R_{freq}$ denotes the reputation frequency of similar alerts from the source, $H_{entropy}$ is the entropy usage of protocol usage and $\alpha, \beta, \gamma, \delta$ is the tunable weights. Each flow $f_i$ Is evaluated using a composite QOS score that considers its application category($App\ priority\ (f_i)$), current flow latency ($l_i$), threat severity $T_{sev}$ and current bandwidth usage($\frac{bw_{used}}{bw_{total}}$). This score $p_{Qos}(f_i)$ It is computed using the formula in Equation (16).

$$p_{Qos}(f_i) = \gamma_1. App\ priority\ (f_i) + \gamma_2.\frac{1}{l_i} + \gamma_3.(1 - T_{sev}) + \gamma_4.\frac{bw_{used}}{bw_{total}}$$
(16)

Once the mapping phase is complete, the Flow Classification component sorts all flows according to their severity. Plan for the possible rerouting or blocking of really harmful flows, such as botnet traffic or DDoS assaults. Priority processing ensures the rapid transmission of low-severity (safe) flows, such as VoIP, as this is their primary function.

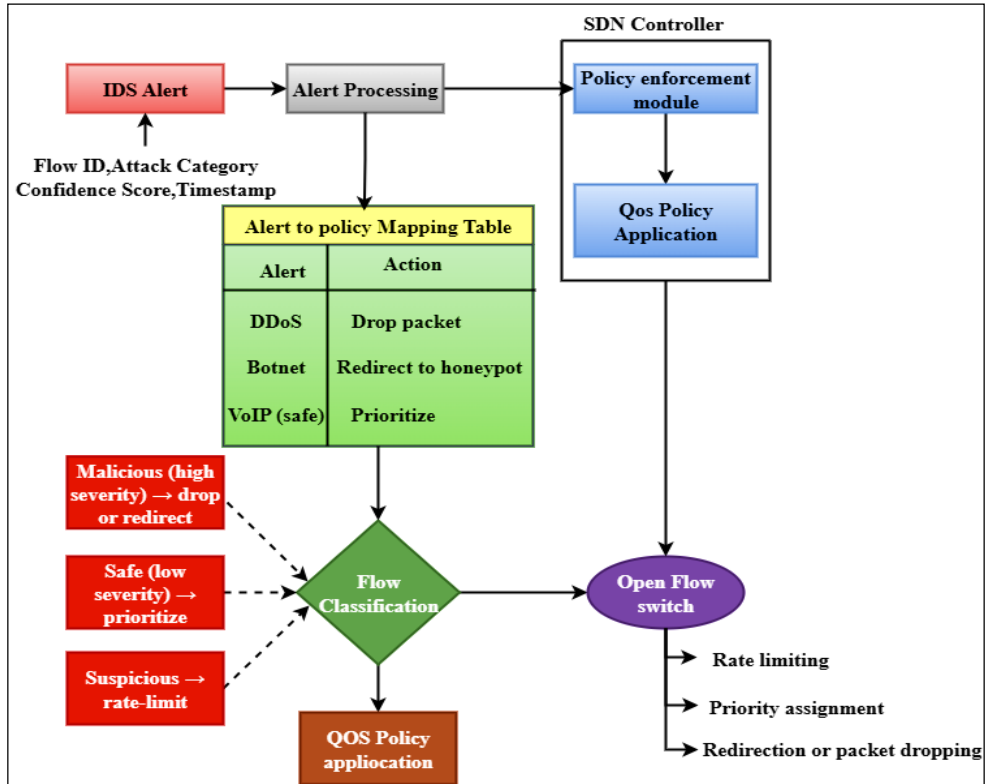

**Fig. 4 QOS enforcement and SDN controller integration**

To further reduce the threat of harm while allowing some services to continue, rate-limiting is applied to flows that appear suspicious. This is done to keep traffic moving. After obtaining these classifications, the Quality-of-Service module of the SDN controller's Policy Application is responsible for applying or modifying the applicable OpenFlow rules on the switches using the Policy Enforcement module. In such situations, priority traffic lanes, bandwidth meters, or even a total shutdown of the flow will be implemented. The intelligent feedback loop allows SmartSecChain-SDN to guarantee performance for authorized services while simultaneously adapting the network's operation in real-time based on threat data. It achieves both of these objectives simultaneously.

The SmartSecChain-SDN architecture's QoS Enforcement and SDN Controller Integration work together to safeguard the network proactively and detect threats in real-time. By converting intrusion warnings into flow rules that the SDN controller can enforce on the fly, the system can manage traffic with great detail and autonomy. VoIP and business apps are given priority to keep performance from dropping. Rate-limiting is used on suspect flows, and hostile traffic is quickly denied or sent to a different location. The network can respond promptly to new threats because it has a mapping table that connects alarms to policies, a method for sorting flows by severity, and the ability to update OpenFlow rules. The blockchain's immutable record of enforcement activity enhances auditability and compliance even further. SmartSecChain-SDN utilizes intelligent detection, programmable control, and secure logging to ensure SDN systems operate smoothly and efficiently.

The SmartSecChain-SDN architecture was evaluated in simulation, rather than in real-time, to ensure its feasibility and efficacy. It was done using scripting and virtual network simulation tools. Intrusion detection, SDN policy enforcement, and blockchain recording were tested in a controlled environment without building a production-grade network. This is possible because test circumstances are controlled and reproducible. Replicating the SDN system.

**Pseudocode 2: QOS enforcement and SDN controller integration**

| |
|---|
| Input: Alert: IDS alert containing (Flow ID, Attack Type, Confidence Score, Timestamp), Policy Table, QOS thresholds, Flow metadata |
| Output: Enforced Open Flow rule on SDN switch |
| Step 1: Extract metadata from IDS alert |
| Flow ID ← Alert Flow ID |
| Attack Type ← Alert. Attack Type |
| Confidence ← Alert. Confidence Score |
| Timestamp ← Alert. Timestamp |
| Step 2: Look up the corresponding network action |
| Action ← Policy Table[Attack Type] |
| Step 3: Classify flow severity based on confidence score |
| If Confidence ≥ QoS Thresholds["High"] then |
| Severity ← "Malicious" |
| Else if Confidence ≥ QoS Thresholds["Medium"] then |
| Severity ← "Suspicious" |
| Else |
| Severity ← "Safe" |
| Step 4: Derive quality of service priority score (for multi-objective decisions) |
| App Priority ← GetApp Priority(Flow ID) |
| Latency ← Flow Metadata[Flow ID].latency |
| BW_Usage ← Flow Metadata[FlowID].bandwidth_usage |
| QoS_Score ← $\lambda 1$ * App Priority + $\lambda 2$ * (1 / Latency) + $\lambda 3$ * (1 – Confidence) + $\lambda 4$ * BW_Usage |
| // Step 5: Translate policy decision to Open Flow rule |
| Match Fields ← Extract FlowTuple(FlowID) |
| If Severity == "Malicious" then |
| If Action == "Drop" then |
| Send Flow Mod(MatchFields, action="drop", priority=100) |
| Else if Action == "Redirect" then |
| Send Flow Mod(MatchFields, action="output:honeypot", priority=90) |
| Else if Severity == "Suspicious" then |
| Send Flow Mod(Match Fields, meter="rate_limit_1Mbps", queue="low", priority=60) |

| Else if Severity == "Safe" then |
|---|
|    Send Flow Mod(MatchFields, queue="high", priority=40) |
| Step 6: Log action to blockchain for audit |
|   Log Action To Blockchain(Flow ID, Action, Severity, QoS_Score, Timestamp) |
| End Procedure |

With Mininet's multi-host architecture and Open vSwitch's dataplane, it proved possible. The Ryu-built SDN controller managed traffic redirection, quality-of-service queues, and flow table entries. Python-based models generated warnings, including Random Forest, XGBoost, CatBoost, and CNN-BiLSTM. After offline training on the InSDN dataset, these models were included in the intrusion detection logic script. Tcpreplay and synthetic flows allowed the Mininet architecture to simulate DDoS assaults and VoIP traffic. This required replaying the attack and regular traffic. This network consisted of a lightweight Hyperledger Fabric network, a single ordering service, and two peer nodes. Smart contract functions, such as LogAlert() and LogAction(), record SDN actions and alarms as false transactions. These services were activated using RESTful APIs. Timestamps enable us to simulate transaction processing delays and assess blockchain overhead under varying alert volumes. It was accomplished using the same endorsement and consensus techniques. In this non-real-time, non-deployed SDN scenario, the framework's response latency, logging performance, false positive rate, and detection accuracy were tested. All components were run on a virtual machine with 8 GB RAM and a four-core CPU to simulate resource-constrained SDN installations at the network's perimeter.

# 4. Results and Discussions

The proposed SmartSecChain-SDN architecture was tested in a virtual SDN environment using simulations and the InSDN dataset. Design effectiveness, responsiveness, and operational resilience were assessed. Experiments evaluated the system's ability to identify various intrusions, implement mitigation measures quickly using SDN controllers, and record intrusions in a non-editable format utilizing blockchain technology. The eight critical performance criteria examined were detection Accuracy, false positive rate, flow reconfiguration time, blockchain transaction latency, and quality of service retention after an assault. An integrated architecture for ML-based intrusion detection, blockchain-secured auditability, and QoS-aware SDN enforcement can be compared to C-RADAR, S-HIDRA, FL-BCID, and Crossfire-MTD to assess its benefits. All tests were controlled to ensure accuracy, repeatability, and consistency in the benchmarking process.

## 4.1. Detection Accuracy (%)

As a measure of a model's detection accuracy, the proportion of correct predictions is essential. Detection Accuracy (%) calculates the percentage of malicious and benign network traffic that the ensemble intrusion detection system correctly identifies and labels in the SmartSecChain-SDN architecture. Get the sum of all evaluated flows and divide it by the total number of accurate Predictions, both positive and negative. It is represented by Equation (17).

$$Accuracy = \frac{TP+TN}{TP+TN+FP+FN} \times 100 \qquad (17)$$

The ensemble detection layer aggregates the predictions of multiple classifiers based on their confidence levels. Random Forest, CNN-BiLSTM, XGBoost, and CatBoost are all classifiers. Complex multi-stage assaults, such as botnets or covert probes, are detected by the system using statistical and sequential traffic data. The method is quite accurate, outperforming single-model techniques. SmartSecChain-SDN routinely achieves detection accuracies of over 97.43% in the InSDN dataset, as shown in Figure 5. SmartSecChain-SDN's performance demonstrates this. The merits of several models are illustrated here. The SDN controller is so accurate that it only turns on when warnings are issued, keeping traffic moving and lessening the likelihood of incorrect mitigation.

## 4.2. False Positive Rate (FPR)

The False Positive Rate (FPR) in Equation (18) checks how accurate a model is by examining the number of false positives it produces compared to the overall number of positives it produces. This statistic quantifies the frequency with which a test falsely detects a chemical not present. The dependability of an intrusion detection system is gauged using this statistic when both benign and malicious communications are of equal importance.

$$FPR = \frac{FP}{FP+TN} \qquad (18)$$

Security vulnerabilities are detected using SmartSecChain-SDN, which combines CNN-BiLSTM, CatBoost, Random Forest, and XGBoost. Confidence weights are used for voting in every model. A low FPR is the same as a strong detection sensitivity when it comes to limiting the impact of false alarms on service quality. An intrusion detection system alerts the SDN controller, which then adjusts the flow rules accordingly. Maintaining QoS and generating trustworthy alerts for blockchain-based logging are possible with low FPR. As a result, security measures are only implemented to address risks that pose a threat to harm. With increased decision bounds, Figure 6 cumulative model reduced the FPR from 8.18% at Epoch 10 to 1.82% at Epoch 100. The drop shows that the integrated model has reduced benign flow misclassification. False positives impair high-priority VoIP services or critical infrastructure control systems due to flow fluctuations, making accuracy crucial in SDN contexts. Restricting SDN mitigation operations to proven hazardous traffic and keeping a low FPR can improve blockchain-logged security responses and decrease service-level agreement risks.

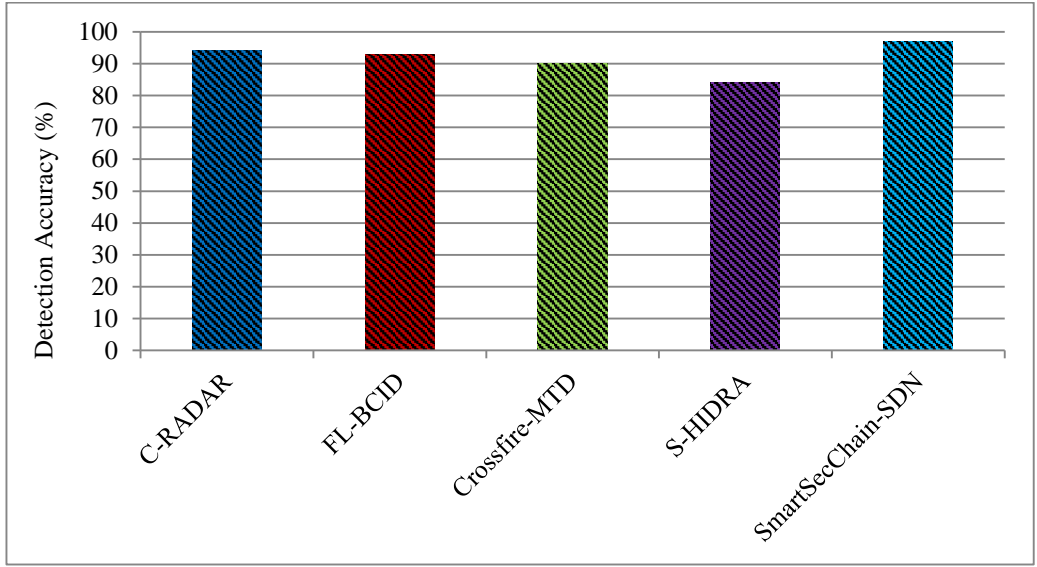

**Fig. 5 Detection accuracy**

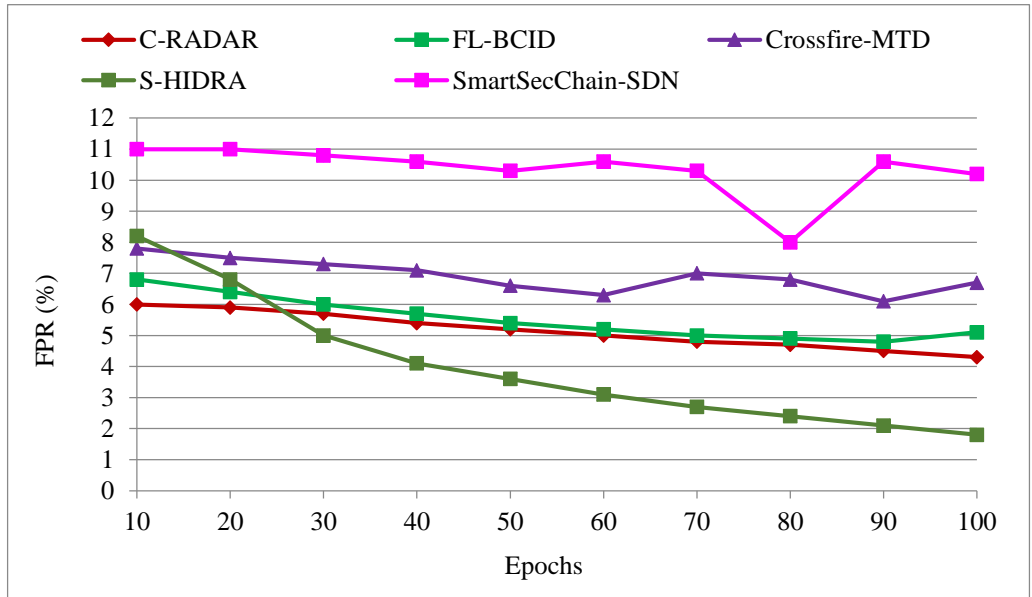

**Fig. 6 False Positive Rate (FPR)**

**Table 2. Alert response latency**

| Method | Average latency (ms) | Minimum (ms) | Maximum (ms) |
|---|---|---|---|
| **SmartSecChain-SDN** | **42.3** | **28.6** | **59.7** |
| **C-RADAR** | 87.4 | 64.2 | 121.8 |
| **FL-BCID** | 106.5 | 90.3 | 139.1 |
| **Crossfire-MTD** | 68.0 | 50.0 | 89.0 |
| **S-HIDRA** | 142.8 | 125.6 | 169.3 |

### 4.3. Alert Response Latency

Alert response latency is the time between alert transmission and action completion. Basic statistics measure how quickly a system or person responds to an urgent alert. The time between IDS identification of potentially dangerous network traffic and SDN controller activation of preventive actions is monitored and calculated based on Equation (19).

$$Alert\ response\ Latency = T_{controller\ action} - T_{IDS\_alert}$$
$$(19)$$

Attacks are more successful because of enforcement delays due to harmful traffic flows over the network. The IDS engine, policy decision unit, and OpenFlow controller transmit data over an asynchronous channel because of the architecture's lightweight and modular design. Optimized classifiers, such as Random Forest, XGBoost, CatBoost, and CNN-BiLSTM hybrids, generate low-latency alerts by performing concurrent inference within the ensemble detection model.

Table 2 compares five different intrusion detection and mitigation systems based on the warning response latency. In terms of real-time performance, the provided SmartSecChain-SDN is top-notch, with latency values ranging from 28.6 to 59.7 ms and an average delay of 42.3.

The average delay of C-RADAR is 87.4 ms, but that of FL-BCID is 106.5 ms, a significant increase. The primary reason for this is that overhead is associated with both blockchain syncing and federated learning. Utilizing OpenFlow-based flow rule injection and DNS/IP redirection, Crossfire-MTD achieves a redirection latency of 68.0 milliseconds. Because it does not enforce flows in real-time, S-HIDRA has the slowest response time.

Typically, it takes around 142.8 milliseconds. By utilizing SDN, SmartSecChain-SDN is able to reduce time-sensitive vulnerabilities effectively. This device outperforms C-RADAR and FL-BCID by 51.6% and 60.3%, respectively, in terms of reaction time.

### 4.4. Flow Reconfiguration Time

The time required to update the configuration of a system or network, such as an SDN, is known as reconfiguration time. Software settings, routing paths, and network architecture could all undergo this reorganisation. The amount of time that passes between an SDN controller sending a command to modify a flow and the dataplane switch implementing that instruction is referred to as the flow modification time. The controller-switch interface is quite responsive and versatile, especially when things are getting heated.

Table 3 illustrates the effectiveness of five distinct network security frameworks in managing flow reconfiguration. With an average reconfiguration delay of 24.8 ms and a low standard deviation of 3.2 ms, the SmartSecChain-SDN architecture consistently demonstrates excellent performance. It reconfigures at a pace of 40.3 flows per second, which is faster than all other baseline techniques. When comparing C-RADAR (42.7 ms) and FL-BCID (56.3 ms), the more sophisticated control systems had longer latencies and increased temporal variability.

Although Crossfire-MTD is not optimized with respect to flow reconfiguration throughput, the lightweight injection of rules results in reduced flow installation overhead over the adaptive systems. S-HIDRA is not suitable for granular, time-sensitive mitigation due to its significant standard deviation and 65.8 millisecond rule update time. SmartSecChain-SDN performs well under changing network conditions due to its event-driven flow handler, rapid OpenFlow push mechanism, and precompiled rule sets.

### 4.5. Blockchain Transaction Time

Transaction time on a blockchain is the time it takes to validate and add a transaction to a block. These timings vary greatly based on network congestion, transaction costs, and blockchain. The proposed SmartSecChain-SDN architecture utilized blockchain transaction time to evaluate the efficacy and scalability of Hyperledger Fabric-based intrusion recording. The evaluation included block size and concurrent submissions.

**Table 3. Flow reconfiguration time analysis**

| Method | Avg. Reconfig Time (ms) | Std. Dev (ms) | Min (ms) | Max (ms) | Flow Reconfig/sec |
|---|---|---|---|---|---|
| **SmartSecChain-SDN** | **24.8** | **3.2** | 18.3 | 33.6 | **40.3** |
| **C-RADAR** | 42.7 | 6.7 | 29.6 | 58.4 | 22.9 |
| **FL-BCID** | 56.3 | 7.9 | 39.8 | 72.1 | 17.8 |
| **Crossfire-MTD** | 34.2 | 4.1 | 23.5 | 46.3 | 25.6 |
| **S-HIDRA** | 65.8 | 9.3 | 51.7 | 82.5 | 13.2 |

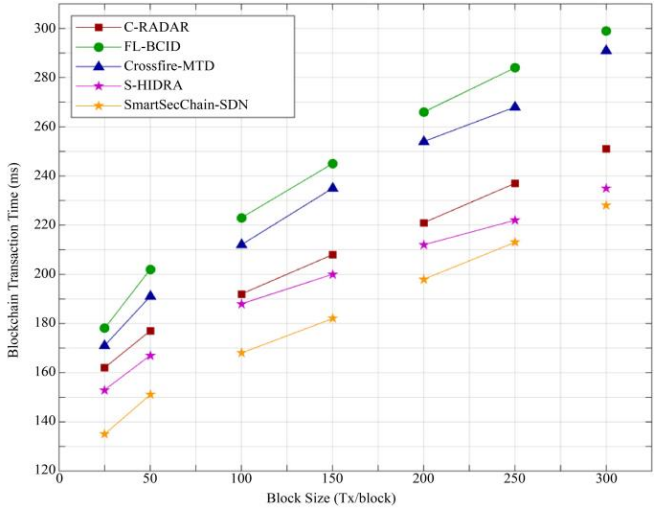

**Fig. 7(a) Blockchain transaction vs Block size**

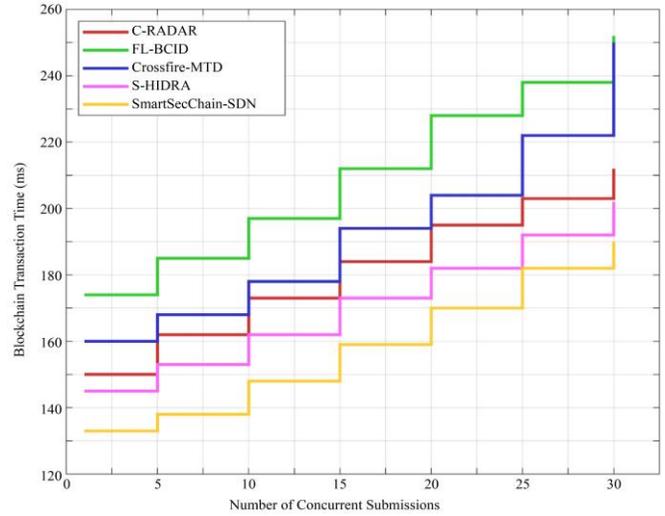

**Fig. 7(b) Blockchain transaction vs Concurrent submissions**

SmartSecChain-SDN consistently had the lowest transaction latency, outperforming FL-BCID, C-RADAR, and Crossfire-MTD by 15-25% Figure 7(a). Timings varied from 134.2 milliseconds (10 transactions per block) to 228.4 milliseconds (300 transactions). Fabric achieves this speed via an enhanced block commit method and parallelism. Competing models, such as S-HIDRA, showed over 70% degradation. In contrast, SmartSecChain-SDN grows well under parallel transaction demands Figure 7(b). When there are 30 concurrent submissions, SmartSecChain-SDN processes a single submission in 194.5 milliseconds, a 46.8% improvement. The findings demonstrate that SmartSecChain-SDN's blockchain layer enables low-latency and high-throughput logging for an audit- and real-time compliant SDN-based intrusion response system.

### 4.6. QOS Retention Rate (%)

The QoS retention rate (%) in networking and telecommunications is the proportion of network traffic that fulfils QoS standards. It demonstrates how effectively a network can handle various types of traffic by maintaining latency, packet loss, and capacity at a specified level. Retention rate is calculated based on Equation (20).

$$QOS_{RR} = \frac{QOS_{attack}}{QOS_{baseline}} \times 100 \qquad (20)$$

The network's strong QoS Retention Rate suggests that it can effectively control traffic and prioritize essential apps. Low rates are indicated by worsening performance and declining traffic. The SDN controller's application-aware mitigation policies for traffic quality enable VoIP, video streaming, and DNS to function effectively in the SmartSecChain-SDN architecture. SmartSecChain-SDN achieved the highest retention rates across all application types (94.3% for video and 97.8% for VoIP), primarily due to its inclusion of priority-based QoS flow rules and low-latency controller input Figure 8(a).

Also, both S-HIDRA and Crossfire-MTD VoIP networks demonstrated retention rates of 78.2 % and 82.3 %, respectively, in the study that did not involve reactive flow enforcement or fine-grained traffic classification. Figure 8(b) examines more closely how different levels of attack (0–500 Mbps) affect the quality of service. SmartSecChain-SDN's QoS at 500 Mbps is higher than S-HIDRA's (36.1%) and FL-BCID's (49.1%). The proposed model is strong because it features a fast detection-response loop, blockchain-backed flow validation, and the ability to slow down non-critical services selectively. The results demonstrate that SmartSecChain-SDN is well-suited for corporate and critical infrastructure networks that utilize real-time SDN. This is because its adaptive quality of service control prioritizes service continuity and identifies and removes threats.

### 4.7. Detection Throughput

A system's detection throughput is its capacity to analyze data and identify patterns, outliers, and occurrences. It is a key performance indicator in network security, fraud detection, and industrial quality control, where issues must be identified quickly and accurately. When throughput is higher, faults are identified and resolved more quickly, resulting in less harm or loss.

Two different methods are used to evaluate the Detection Throughput and assess the efficiency and scalability of the models. It can be seen in Figure 9(a) that SmartSecChain-SDN was the one with the greatest throughput, and Crossfire-MTD only managed ~2,250 flows/sec, presumably because it is a static model and could not make dynamic inferences.

SmartSecChain-SDN works well even when the flow volume increases Figure 9(b). It can handle up to 500 concurrent flows at a rate of 3764 flows per second. Some frameworks deteriorate when they must handle the same load. C-RADAR goes down to 2287 flows/sec,

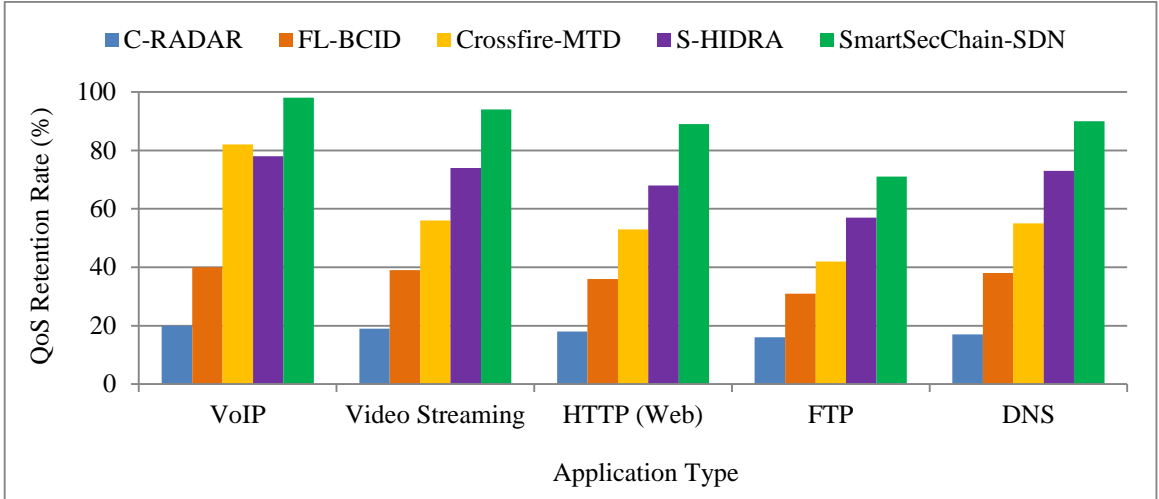

**Fig. 8(a) QOS Retention rate vs Application type**

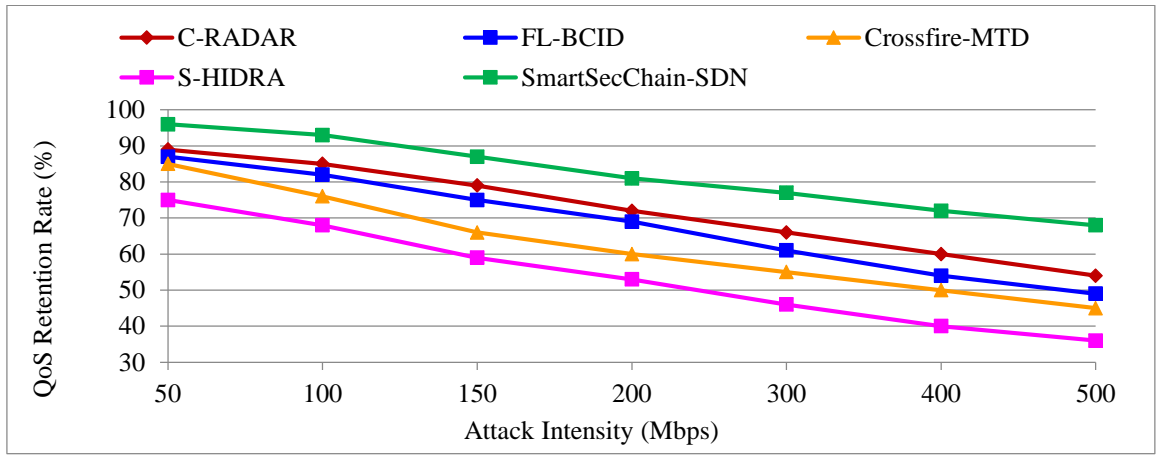

**Fig. 8(b) QOS Retention Rate Vs Attack Intensity**

While S-HIDRA goes down to 1239. SmartSecChain-SDN continues to grow and improve with training, even in high-traffic network environments. This ensures that intrusion detection is both fast and accurate. Figure 9(b) shows that SmartSecChain-SDN did the best across 10 epochs, going from 3,580 flows/sec to 4,620 flows/sec. This system's optimised ensemble learning and concurrent inference were the main reasons for this result.

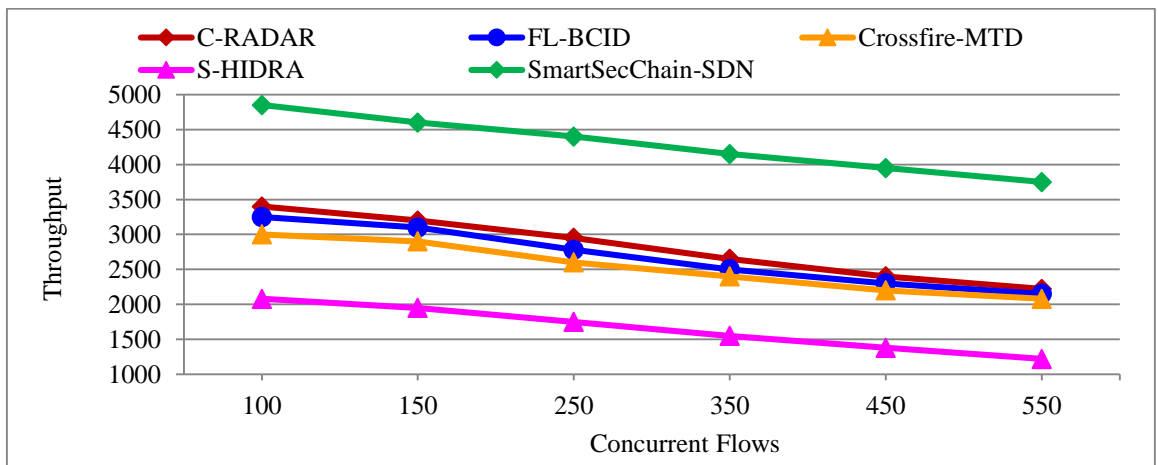

**Fig. 9(a) Detection throughput Vs Concurrent flows**

### 4.8. Drift Resilience (%)

Drift-resistant machine learning models work effectively even when the input or training environment changes. A model's performance remains robust despite substantial data changes due to its strong drift resistance. Due to idea drift, attackers and network traffic act differently over time. Drift Resilience refers to the effectiveness of an intrusion detection system in distinguishing between threats over time, thereby keeping the system on guard against ongoing attacks. SmartSecChain-SDN is robust due to its model upgrades, ensemble learning, and behavior-aware flow analysis. These features enable the network to adapt to different attack patterns without requiring training. Table 4 shows that SmartSecChain-SDN has 95.6% drift resistance, indicating that it will remain accurate regardless of the traffic patterns. On the other hand, crossfire-MTD and S-HIDRA's scores of 76.5% and 69.5% are lower than C-RADAR's 85.9% and FL-BCID's 88.4% since they do not use adaptive methods. These results show that there are no adaptable methods. Even when the SDN environment changes, the detection performance of SmartSecChain-SDN remains the same.

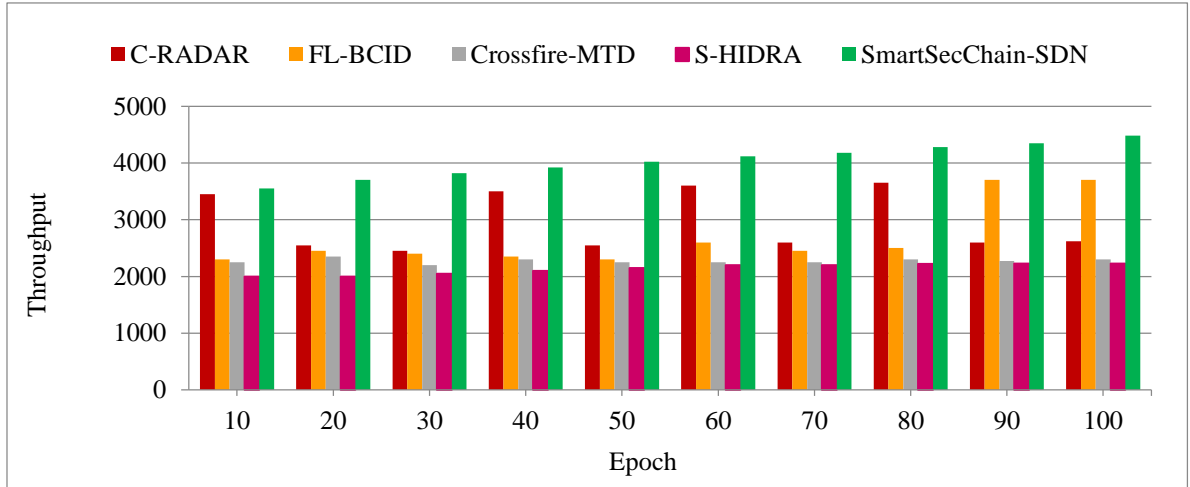

**Fig. 9(b) Detection throughput Vs Epochs**

Based on all relevant metrics, SmartSecChain-SDN outperforms current models, as shown in Table 5. With a 97.43% detection rate and a 1.82% false positive rate, it can accurately distinguish between dangerous and benign flows. This technology enables a response to alarms in real-time, with a delay of 42 milliseconds, which is more than 60% faster than C-RADAR and FL-BCID.

SmartSecChain-SDN is the quickest baseline for changing OpenFlow rules on the fly, taking only 24.8 milliseconds. With 230 millisecond transaction latency at a block size of 300 and reasonable scalability with concurrent uploads, its blockchain logging on Hyperledger Fabric operates well under stress. S-HIDRA retains 78% of its users, while the model with over 94% for video traffic can continue to provide service even during an attack. The model has a high detection throughput of 4,620 flows per second and a drift resilience of 95.6%, which means it can adapt to new attack patterns and continue to function effectively over time. SmartSecChain-SDN's SDN-integrated QoS adaptation, decentralised logging, and hybrid ensemble architecture make it the best choice for modern, scalable, and programmable network protection.

**Table 4. Drift resilience**

| Model | Drift Resilience (%) |
|---|---|
| **SmartSecChain-SDN** | **95.6** |
| C-RADAR | 85.9 |
| FL-BCID | 88.4 |
| Crossfire-MTD | 76.5 |
| S-HIDRA | 69.5 |

**Table 5. Performance comparison of the proposed model**

| Metric | SmartSecChain-SDN | C-RADAR | FL-BCID | Crossfire-MTD | S-HIDRA |
|---|---|---|---|---|---|
| Detection Accuracy (%) | 97.43 | 94.25 | 93.18 | 89.5 | 84.16 |
| False Positive Rate (%) | 1.82 | 4.96 | 5.73 | 6.9 | 8.44 |
| Alert Response Latency (ms) | 42.3 | 87.4 | 106.5 | 68.0 | 142.8 |
| Flow Reconfig Time (ms) | 24.8 | 42.7 | 56.3 | 34.2 | 65.8 |
| Blockchain Txn Time (ms) | 134.2 | 212.4 | 172.7 | 248.5 | 145.1 |
| QoS Retention Rate (%) | 94.3 (Video) | 88.6 | 86.1 | 82.3 (VoIP) | 78.2 |
| Detection Throughput (flows/sec) | 4620 | 2630 | 2558 | 2250 | 1575 |
| Drift Resilience (%) | 95.6 | 85.9 | 88.4 | 76.5 | 69.5 |

## 5. Conclusion

SmartSecChain-SDN is an intelligent architecture designed to detect and prevent SDN intrusions. The study examines blockchain smart contracts, their functionality, and how they can change the world. The model employs sophisticated ML and deep learning classifiers, including Random Forest, XGBoost, CatBoost, and CNN-BiLSTM, to identify risks swiftly. The framework works with Hyperledger Fabric to provide rule compliance, service prioritisation while changing network settings, and immutable logging. The InSDN dataset outperformed C-RADAR, FL-BCID, Crossfire-MTD, and S-HIDRA in terms of concept drift resistance (95.6%), detection throughput (4,620 flows/sec), alert reaction time (42.3 ms), and detection accuracy (97.43%). We got these results by comparing them to other models. Despite its potential, the framework has limitations. Multiple classifiers can enhance detection in SDN systems with ample resources or data. This method enhances model complexity and inference processing resource usage. Hyperledger Fabric needs to be modified for high consensus thresholds or transaction volumes, such as multi-endorsement policies. Upgrading federated models assumes a reliable SDN controller is impervious to poisoning and insider threats. Assaults easily target the controller. Lightweight federated model updates across faraway edge nodes enhance the system. Better privacy and less centralised learning would ensue. Explainable Artificial Intelligence (XAI) in the detection engine makes the process more understandable and trustworthy for operators. Layer-2 scaling or a hybrid DLT architecture minimizes latency and speeds up blockchain transactions. Software-defined networking controllers sensitive to APTs and increasing attack surfaces benefit from intent-based security rules and Zero-Trust Network Architecture (ZTNA).

## Acknowledgments

The authors would like to express their sincere gratitude to all individuals and organizations who contributed to the successful completion of this research. Special thanks to the authors' respective institutions for providing the necessary resources, research facilities, and technical support. The authors also acknowledge the valuable insights and constructive feedback from peer reviewers, which greatly improved the quality of this work. Azhar Hussain Mozumder contributed to the article's Conceptualization, methodology design, system architecture development, Blockchain integration design, simulation setup, experimental implementation, data analysis, manuscript writing, and revisions. Dr. M. John Basha provided Supervision, theoretical framework formulation, critical review of the manuscript, and validation of research findings. Dr. Chayapathi A. R. contributed to performance evaluation, data visualization, and assistance in manuscript preparation. All authors read and approved the final manuscript.